\documentclass[sigconf]{acmart}
\renewcommand\footnotetextcopyrightpermission[1]{} 

\usepackage{textcomp}
\usepackage{listings}
\usepackage{xcolor}
\usepackage{graphicx}
\usepackage{graphics}
\usepackage{subcaption}
\usepackage{balance}
\usepackage[ruled,linesnumbered]{algorithm2e}
\usepackage{multirow}
\usepackage{makecell}
\usepackage{float}
\usepackage{hyperref}

\usepackage{subcaption}
\definecolor{mygreen}{rgb}{0,0.6,0}
\definecolor{mygray}{rgb}{0.5,0.5,0.5}
\definecolor{mymauve}{rgb}{0.58,0,0.82}

\begin{document}
\pagestyle{plain} 
\title{Are LLMs Any Good for High-Level Synthesis?}

\author{Yuchao Liao}
\affiliation{
  \institution{Electrical and Computer Engineering\\University of Arizona}
  \city{Tucson}
  \country{Arizona, USA}}
\email{yuchaoliao@arizona.edu}

\author{Tosiron Adegbija}
\affiliation{
  \institution{Electrical and Computer Engineering\\University of Arizona}
  \city{Tucson}
  \country{Arizona, USA}}
\email{tosiron@arizona.edu}

\author{Roman Lysecky}
\affiliation{
  \institution{Electrical and Computer Engineering\\University of Arizona}
  \city{Tucson}
  \country{Arizona, USA}}
\email{rlysecky@arizona.edu}

\renewcommand{\shortauthors}{Liao et al.}

	
\begin{abstract}

The increasing complexity and demand for faster, energy-efficient hardware designs necessitate innovative High-Level Synthesis (HLS) methodologies. This paper explores the potential of Large Language Models (LLMs) to streamline or replace the HLS process, leveraging their ability to understand natural language specifications and refactor code. We survey the current research and conduct experiments comparing Verilog designs generated by a standard HLS tool (Vitis HLS) with those produced by LLMs translating C code or natural language specifications. Our evaluation focuses on quantifying the impact on performance, power, and resource utilization, providing an assessment of the efficiency of LLM-based approaches. This study aims to illuminate the role of LLMs in HLS, identifying promising directions for optimized hardware design in applications such as AI acceleration, embedded systems, and high-performance computing.

\end{abstract}


\maketitle

\section{Introduction}
The increasing demand for custom hardware accelerators, driven by applications ranging from artificial intelligence to high-performance computing, necessitates innovative design methodologies to meet the challenges of rapidly evolving technology. High-Level Synthesis (HLS) has emerged as a valuable approach for designing, synthesizing, and optimizing hardware systems. HLS \cite{coussy09introduction} enables designers to define systems at a high abstraction level, independent of low-level circuit specifics, and utilize HLS tools to produce an optimized low-level hardware description of the target system. With current HLS tools (e.g., Vitis HLS, SmartHLS), designers can create application-specific embedded systems using high-level languages like C/C++ and translate them into register-transfer level (RTL) implementations using hardware description languages (e.g., Verilog. VHDL), thereby enhancing design productivity and reducing both design time and cost. Despite the advantages of HLS, the tools can still be time-consuming to use and demand considerable expertise, thus creating the potential for substantial improvement, especially with the integration of technologies like large language models (LLMs).

Recent advancements in LLMs \cite{zhao2023_llm_survey} have showcased their ability to automate various computational tasks, including code generation and software engineering. This presents a unique opportunity to explore the potential of LLMs in streamlining the HLS process, from high-level language specifications to efficient hardware implementations \cite{chang2023_chipgpt}. The ability of LLMs to understand and generate code, combined with the potential for natural language interaction, can revolutionize the way we design hardware, making the process more accessible and less time-consuming. This integration can lead to significant improvements in design productivity and efficiency, ultimately transforming the landscape of hardware development.

In this paper, we explore the burgeoning field of LLMs for HLS, which has sparked growing interest. We first present a taxonomy of LLM use cases for HLS, highlighting the various ways these models can be integrated into the design flow. Building on this foundation, we survey the state-of-the-art, highlighting the most promising research and techniques. To assess the viability of LLMs in the HLS design flow, we perform an experimental evaluation, comparing the Verilog designs generated using a standard HLS tool, specifically Vitis HLS, to those produced with LLM-based approaches. These approaches include direct LLM translation of C benchmarks from the PolyBench Suite \cite{Polybench} to Verilog using ChatGPT-4o, and the use of LLMs to interpret natural language specifications into both benchmarks and Verilog. Our evaluation focuses on the quality (performance, power, resource utilization) of designs produced by each methodology.

This study seeks to answer several key questions: Can existing LLMs generate Verilog code comparable in quality to that produced by traditional HLS tools? What are the advantages and limitations of using LLMs in this context? Could the natural language understanding capabilities of LLMs open up new avenues for hardware design? By addressing these questions, we aim to provide valuable insights into the role of LLMs in HLS and their potential to transform the future of hardware design.

\begin{figure}[t]
    \centering
    \includegraphics[width=0.6\columnwidth,keepaspectratio]{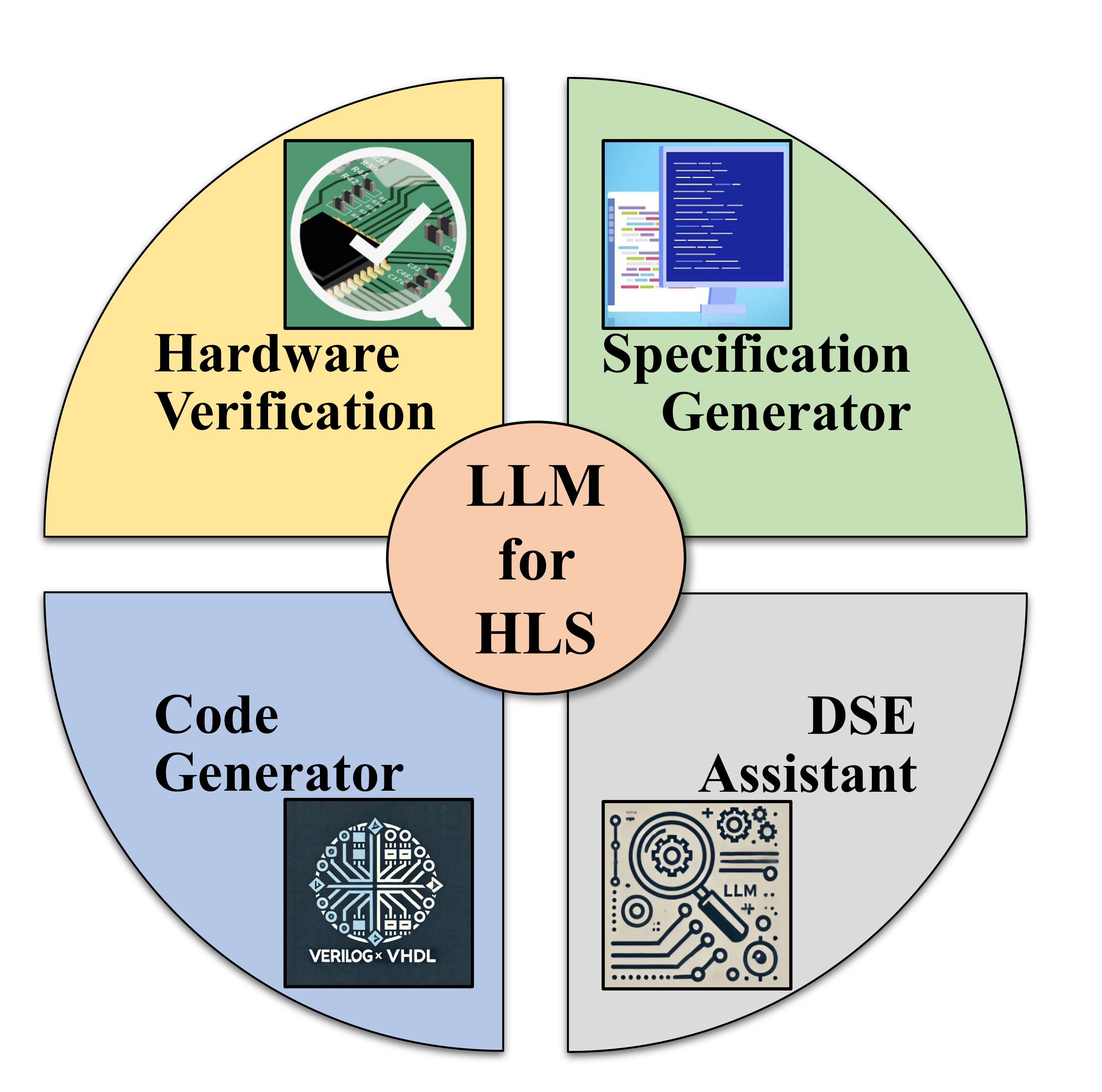}
    \vspace{-10pt}
    \caption{Taxonomy of LLM applications in HLS}
    \label{fig:survey_chart}
\end{figure}

\section{Taxonomy of LLM for HLS}

The application of LLMs to different stages of the HLS process has emerged as a promising research direction. To provide a structured overview of this evolving landscape, we present a taxonomy (illustrated in Figure \ref{fig:survey_chart}) that categorizes LLMs based on their primary role in HLS: \textit{specification generators}, \textit{design space exploration assistants}, \textit{code generators}, and \textit{hardware verification} tools. This classification provides a framework for understanding how LLMs can augment HLS methodologies, as detailed in the following subsections.

\subsection{LLM as Specification Generator}

LLMs hold promise as specification generators in HLS, translating natural language or higher-level code into HLS-compatible formats (e.g., HLS-C) \cite{swaroopa2024_evaluatingLLM,collini2024_c2hlsc,xu2024_programRepair}. This allows for intuitive and accessible expression of hardware functionality. Challenges persist in mitigating ambiguities inherent in natural language, which can lead to misinterpretations. Techniques like prompting, clarification dialogues, and formal verification are crucial for ensuring the correctness of LLM-generated specifications \cite{lu2024rtllm}.

\subsection{LLM as Code Generator}

LLMs can help with code generation, directly generating synthesizable HDL from high-level specifications \cite{blocklove2023_chipChat,thakur2024verigen,chang2023_chipgpt}. This automation can boost productivity and reduce errors. The challenge lies in ensuring generated code quality and providing designers control over code structure and style \cite{lu2024rtllm}.  Recent research demonstrates LLM capabilities in generating functional HDL for various hardware components, including arithmetic units \cite{liu2023verilogeval}, controllers, and simple processors \cite{blocklove2023_chipChat}, suggesting a promising future for this approach. 

\subsection{LLM as Hardware Verification Assistant}

LLMs can assist with hardware verification in HLS, by automating the generation of test cases and identifying potential design flaws \cite{ahmad2024_hardware_bug_repair, kande2023_llm_assertion}. This can lead to significant time savings and improved design reliability. However, challenges persist in ensuring the accuracy of LLM-generated test cases and their integration into existing HLS workflows. Ongoing research \cite{orenesvera2023_usingllmsfacilitateformal} explores the potential of LLMs in areas like formal verification, further highlighting their potential in ensuring the correctness of complex designs.

\subsection{LLM as Design Space Exploration Assistant}

Although receiving less attention than other applications, LLMs are promising in aiding HLS design space exploration (DSE) by suggesting optimizations and exploring design alternatives \cite{liao2023efficient}. Their ability to analyze design constraints and objectives can lead to faster design cycles and innovative solutions. However, effective LLM DSE assistance requires incorporating domain-specific knowledge and addressing potential biases in suggestions. Recent research shows LLMs can optimize hardware accelerators, explore neural network architectures, and propose circuit-level optimizations, emphasizing their transformative potential for DSE \cite{thakur2023_llm_rtl}.

\section{Survey of the State-of-the-Art in LLMs for HLS}

This section surveys the diverse applications of LLMs in HLS, spanning hardware design automation, software-hardware co-design, and design of embedded systems. We examine key research areas such as natural language processing (NLP) to HDL translation, code generation, optimization and verification, and multimodal approaches. We also discuss input modalities used in the state-of-the-art, like textual descriptions and pseudocode, and the output modalities such as HDLs (VHDL, Verilog, SystemVerilog) and HLS-compatible programs (e.g., HLS-C). Finally, we highlight current approaches to benchmarking and evaluating LLM-driven HLS, emphasizing the need for standardized metrics and datasets to facilitate fair comparisons and drive further advancements in this rapidly evolving field.

\subsection{LLMs Used for HLS}

Recent advancements in LLMs such as ChatGPT, Gemini, Claude, and LLAMA have great potential for use in HLS. While many current works leverage the popular ChatGPT for their HLS experimentation, both general-purpose and custom-tuned LLMs have been utilized to automate and optimize synthesis processes \cite{fu2024_hardwarephi15blargelanguage}. As expected, fine-tuning models on domain-specific data often yields superior performance in generating desired outputs within the HLS workflow. For instance, Nadim et al. \cite{nadimi2024_llmVerilogGen} introduced a multi-expert LLM architecture to address the challenges of design complexity. By using specialized models and a complexity classifier, they achieved an improvement of up to 23.9\% in the pass@k metric. However, a consistent theme emerging from both existing literature and our experiments is the necessity of human-in-the-loop (HITL) approaches for successful LLM integration in HLS. For example, Collini et al. \cite{collini2024_c2hlsc} highlighted the significant human expert guidance required for converting a C-based QuickSort kernel to HLS-C. Similarly, Swaroopa et al. \cite{swaroopa2024_evaluatingLLM} demonstrated a semi-automated approach for generating HLS-C from natural language using LLMs, acknowledging the need for human intervention in the design process, though their work did not evaluate the quality of the resulting designs. Such a HITL approach leverages the computational strengths of LLMs while retaining the nuanced understanding and decision-making capabilities of human experts, to achieve superior HLS outcomes.

\subsection{Applications}

The increasing interest in applying LLMs to HLS has led to promising developments across various domains. For example, LLMs have shown success in automating the generation of analog/mixed-signal (AMS) circuit netlists from transistor-level schematics \cite{tao2024amsnet}. In the domain of RTL generation, LLMs have demonstrated their capability to generate RTL code from natural language descriptions \cite{lu2024rtllm} and, as explored in \cite{blocklove2023_chipChat}, have the potential to aid in writing and debugging HDL code through conversational interactions with existing LLM tools like ChatGPT. Additionally, LLMs are being integrated into tools like MATLAB and Simulink to translate high-level design specifications into synthesizable Verilog and VHDL code, streamlining the HDL generation process. In the domain of code security, Nair et al. \cite{nair2023_generatingsecurehardware} investigated the vulnerabilities in hardware code generated by ChatGPT, specifically analyzing common weaknesses enumerations (CWE) and proposing strategies to guide secure hardware code generation.

Beyond these applications, LLMs are being explored for broader roles in the HLS workflow. Recent work has explored the potential of LLMs to refactor existing C code into HLS-compatible formats, bridging the gap between software and hardware design \cite{collini2024_c2hlsc,fu2023_gpt4aigchip,swaroopa2024_evaluatingLLM, xu2024_programRepair}. Models like ChatGPT have been leveraged to convert high-level design specifications into synthesizable HDL, targeting specific hardware components such as random number generators \cite{meech2023leveragingLLM}. They have been used for automated code repair and optimization to improve the quality of HLS-C programs \cite{xu2024_programRepair}. Furthermore, LLMs have shown promise in generating HLS pragmas \cite{fu2023_gpt4aigchip, xu2024_programRepair}, which are compiler directives that can significantly impact the quality of the generated hardware. Moreover, the use of LLMs for automated testbench generation \cite{qiu2024autobench,bhandari2024_testbench} and hardware design verification tasks \cite{ahmad2024_hardware_bug_repair,kande2023_llm_assertion} further expands their potential applications in HLS. The growing breadth of LLM applications in HLS underscores their potential to enhance automation, efficiency, and accessibility throughout the hardware design process.

\begin{figure*}[t]
    \centering
    \begin{subfigure}[b]{1\textwidth}
        \centering
        \includegraphics[width=0.75\textwidth,height=\textheight,keepaspectratio]{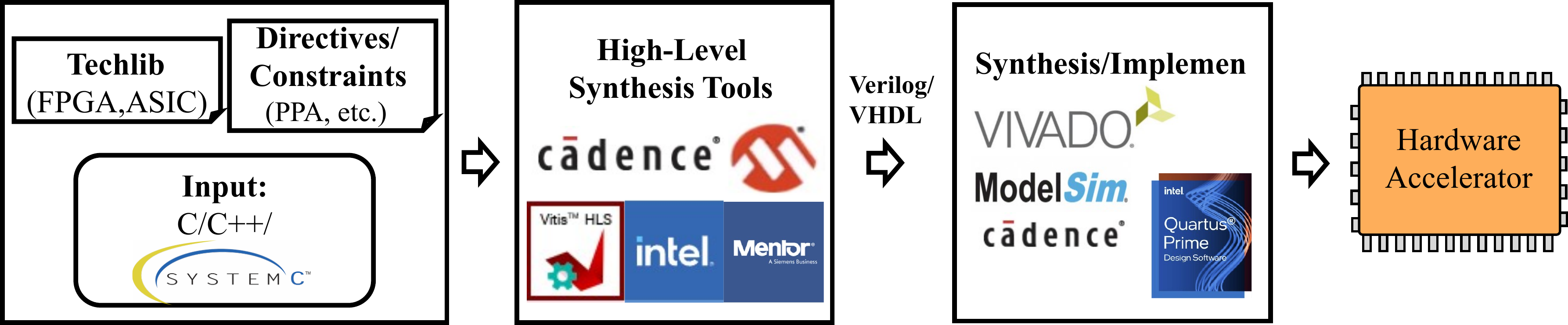}
        \caption{HLS-based approach}
        \label{fig:HLS_method}
    \end{subfigure}
    \par\bigskip
    \begin{subfigure}[b]{1\textwidth}
        \centering
        \includegraphics[width=0.75\textwidth,height=\textheight,keepaspectratio]{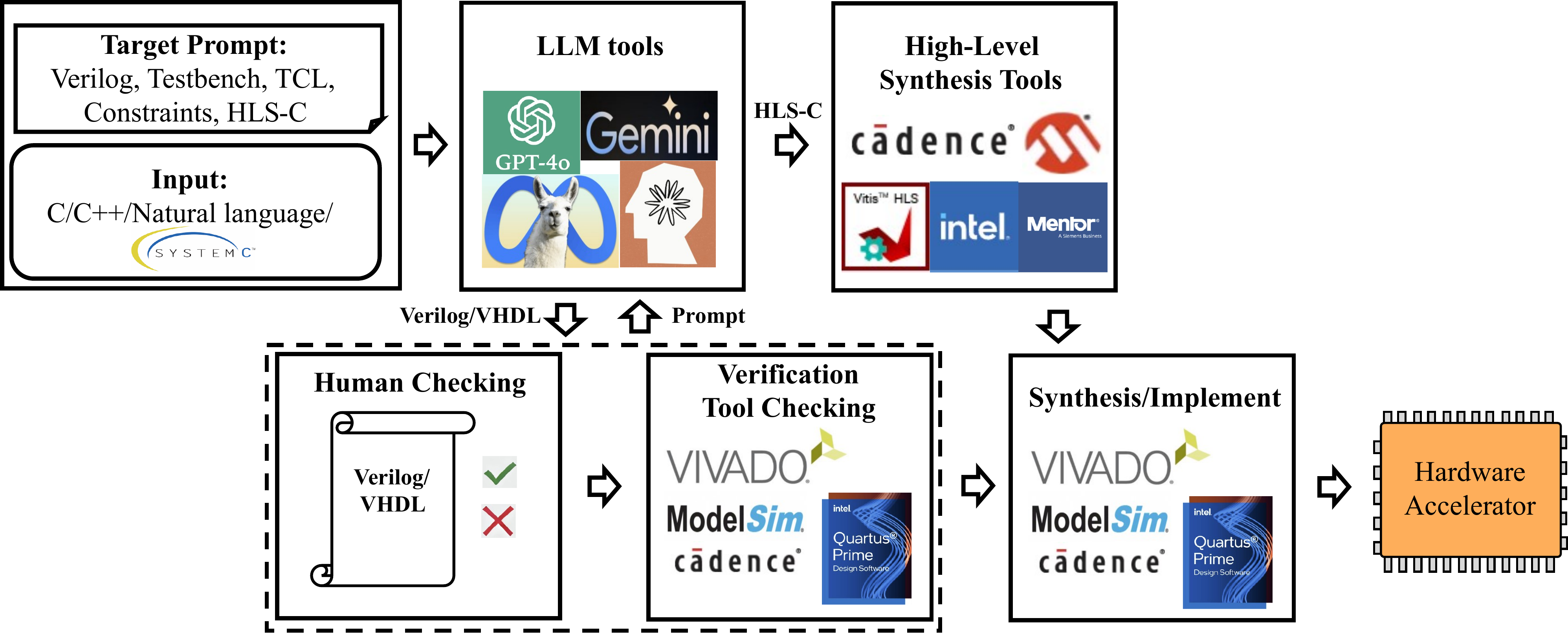}
        \caption{LLM-based approaches}
        \label{fig:LLM_method}
    \end{subfigure}
    \vspace{-10pt}
    \caption{HLS-based (a) and LLM-based (b) approaches to generating hardware accelerators}
    \label{fig:methods}
\end{figure*}

\subsection{Input and Output Modalities}

The versatility of LLMs in HLS stems, in part, from their ability to process and generate information across diverse modalities. Textual descriptions, including high-level design specifications, natural language explanations of functionality, and code snippets in languages like C/C++ often serve as primary input modalities. LLMs can transform these textual inputs into HDL such as Verilog or VHDL, as seen in applications that convert natural language descriptions directly to HDL \cite{meech2023leveragingLLM,blocklove2023_chipChat,lu2024rtllm}. Beyond text, advanced LLMs are increasingly capable of handling multimodal inputs, which incorporate images, schematics, or other data types \cite{chang2024_multi_modal}. This can allow for a more nuanced understanding of design requirements by integrating visual and textual information.

The output modalities of LLMs for HLS are equally diverse. Primarily, LLMs can generate synthesizable HDL code from textual or multimodal inputs \cite{lu2024rtllm}. Additionally, LLMs can optimize existing code by automatically inserting and tuning pragmas to enhance the synthesis process. Moreover, LLMs can generate testbenches and verification scripts, which are vital to validate the functionality and performance of the synthesized hardware. 

\subsection{Benchmarking and Evaluation}

The evaluation and advancement of LLMs in HLS rely on robust benchmarks and datasets. Several key initiatives have emerged to address this need, including the RTLLM benchmark \cite{lu2024rtllm}, which provides a framework for evaluating LLM performance in generating RTL from natural language instructions, encompassing syntax, functionality, and code quality. The RTL-Repo benchmark \cite{allam2024rtl_repo} expands this evaluation by assessing LLM capabilities in generating Verilog code autocompletions within large-scale and complex RTL projects, reflecting real-world design scenarios. VerilogEval \cite{liu2023verilogeval} is a framework for evaluating the effectiveness of LLMs in generating Verilog code, including tasks like module implementation, code debugging, and testbench construction, to assess their potential in hardware design automation. Similarly, VHDL-Eval \cite{vijayaraghavan2024_vhdl_eval} is a specialized framework designed to evaluate LLM performance specifically in VHDL code generation. Wan et al. \cite{wan2024_buginsert_dataset} explored using LLMs to insert bugs into HLS code, and created a dataset including both correct and injected buggy codes. These benchmarks and datasets, along with other emerging efforts, are crucial in LLM-driven HLS research, facilitating the evaluation of LLM capabilities and guiding the development of more robust HLS solutions.

\section{Experimental Methodology}

This section details our experimental methodology for evaluating the effectiveness of integrating LLMs into the HLS process. We aim to assess both the design process and the quality of the hardware generated using LLMs in comparison to solely using traditional HLS tools. We investigate four approaches: 
\vspace{-5pt}
\begin{enumerate}
    \item \textit{Baseline:} Generating Verilog using a standard HLS tool (Vitis HLS) from C code.
    \item \textit{Direct LLM translation:} Employing LLMs to translate C code into Verilog.
    \item \textit{Natural language to Verilog:} Directly generating Verilog code from natural language specifications using LLMs.
    \item \textit{Natural language to code:} Using LLMs to interpret natural language specifications into HLS-C benchmarks, which are then translated into Verilog using Vitis HLS.
\end{enumerate}

\subsection{HLS Approach}
The general HLS design flow, as illustrated in Figure \ref{fig:HLS_method}, transforms a high-level language input to a synthesizable hardware description (e.g., in Verilog or VHDL). This process starts with describing the desired hardware functionality in a high-level language like C/C++/SystemC), followed by synthesis for a specific hardware target, e.g., FPGAs like the Artix-7 or Zynq UltraScale+. We refer to this process as \textit{C$\rightarrow$HLS$\rightarrow$Verilog}.

HLS tools offer a range of directives to guide the synthesis process, allowing designers to control various aspects of the design, such as loop unrolling, pipelining, array partitioning, and performance optimization. While these directives provide flexibility, the resulting HDL code generated by HLS tools can often be complex and challenging to interpret for designers who are primarily accustomed to higher-level programming languages. This limited visibility into the generated HDL code is a key consideration that motivates the exploration of LLMs in HLS, aiming to improve the design process by providing higher-level abstractions or enhancing code understandability.

\begin{figure}
\centering
\begin{subfigure}[b]{0.95\linewidth}
\begin{lstlisting}[]
Background: I am a hardware engineer, and I want to transfer a C program to synthesizable Hardware Description Language (RTL) in Verilog with a full module, including I/O and functionalities from the C program. This Verilog module should target a Xilinx FPGA board.
\end{lstlisting}
\vspace{-10pt}
\caption{Prompt}
\label{fig:prompt_background}
\end{subfigure}

\begin{subfigure}[b]{0.95\linewidth}
\begin{lstlisting}[]
Sure, I can help with that. Please provide the C program you want to translate to Verilog. Additionally, let me know if there are any specific functionalities or constraints you need to consider for the translation, such as the I/O pin count or specific features of the Xilinx FPGA board you are targeting.
\end{lstlisting}
\vspace{-10pt}
\caption{Feedback}
\label{fig:feedback_background}
\end{subfigure}
\vspace{-10pt}
\caption{Background, role, and constraints}
\label{fig:LLM_background}
\end{figure}

\begin{figure}
\centering
\begin{subfigure}[b]{0.95\linewidth}
\begin{lstlisting}[language=C]
#define N 64

void syrk(float alpha, float beta, 
    float A[N][N], float B[N][N], float C_out[N][N]) {
    int i, j, k;

    float buff_A0[N][N];
    float buff_B1[N][N];
    float buff_B[N][N];
    float buff_C_out[N][N];
\end{lstlisting}
\vspace{-10pt}
\caption{Prompt}
\label{fig:prompt_IO1}
\end{subfigure}

\begin{subfigure}[b]{0.95\linewidth}
\begin{lstlisting}[language=verilog]
`timescale 1ns / 1ps
module syrk#(parameter N = 64) (
    input [31:0] alpha, 
    input [31:0] beta,  
    integer i, j, k;
    input [31:0] A[0:N-1][0:N-1], 
    input [31:0] B[0:N-1][0:N-1], 
    output reg [31:0] C_out[0:N-1][0:N-1]);
    reg [31:0] buff_A0[0:N-1][0:N-1];
    reg [31:0] buff_A1[0:N-1][0:N-1];
    reg [31:0] buff_B[0:N-1][0:N-1];
    reg [31:0] buff_C_out[0:N-1][0:N-1];);
\end{lstlisting}
\vspace{-10pt}
\caption{Feedback}
\label{fig:feedback_IO1}
\end{subfigure}

\begin{subfigure}[b]{0.95\linewidth}
\begin{lstlisting}[language=C]
The inputs A, B, and C_out are not correct. We need an external RAM to handle the 2D array. The width of the address and the width of the data should be declared as module parameters.
\end{lstlisting}
\vspace{-10pt}
\caption{Prompt2}
\label{fig:prompt_IO2}
\end{subfigure}

\begin{subfigure}[b]{0.95\linewidth}
\begin{lstlisting}[language=verilog]
module syrk#(parameter N = 64,DATA_WIDTH = 32)(
    input clk,
    input rst,
    input [DATA_WIDTH-1:0] alpha,
    input [DATA_WIDTH-1:0] beta,
    input [DATA_WIDTH-1:0] A_data,
    input [DATA_WIDTH-1:0] B_data,
    output reg [DATA_WIDTH-1:0] C_out_data,
    input [$clog2(N*N)-1:0] A_addr, B_addr, 
            C_out_addr,
    input A_we, B_we, C_out_we);
\end{lstlisting}
\vspace{-10pt}
\caption{Feedback2}
\label{fig:feedback_IO2}
\end{subfigure}
\vspace{-5pt}
\caption{Inputs and error checking (I/O)}
\label{fig:LLM_IO}
\end{figure}

\begin{figure}
\centering
\begin{subfigure}[b]{0.95\linewidth}
\begin{lstlisting}[language=C]
...
lp1: for (i = 0; i < N; i++) {
    lp2: for (j = 0; j < N; j++) {
        if (j > i) continue;
        lp3: for (k = 0; k < N; k++) {
              buff_C_out[i][j] += alpha * 
              buff_A0[i][k] * buff_A1[j][k];
        }
    }
}
...
\end{lstlisting}
\vspace{-10pt}
\caption{Prompt}
\label{fig:prompt_loop1}
\end{subfigure}


\begin{subfigure}[b]{0.95\linewidth}
\begin{lstlisting}[language=verilog]
...
for (i = 0; i < N; i = i + 1) begin 
    for (j = 0; j < N; j = j + 1) begin 
        if (j <= i) begin 
            for (k = 0; k < N; k = k + 1) 
            begin        
                buff_C_out[i][j] <= buff_C_out[i][j] 
                + alpha * buff_A0[i][k] 
                * buff_A1[j][k]; 
            end
            buff_C_out[i][j] <= buff_C_out[i][j] 
            + beta * buff_B[i][j]; 
        end else begin 
            C_out[i][j] <= 0; 
        end 
    end 
end
...
\end{lstlisting}
\vspace{-10pt}
\caption{Feedback}
\label{fig:feedback_loop1}
\end{subfigure}

\begin{subfigure}[b]{0.95\linewidth}
\begin{lstlisting}[language=C]
This looks wrong, A, B, C_out, are external BRAM, please update this again.
\end{lstlisting}
\vspace{-10pt}
\caption{Prompt2}
\label{fig:prompt_loop2}
\end{subfigure}

\begin{subfigure}[b]{0.95\linewidth}
\begin{lstlisting}[language=verilog]
...
COMPUTE: begin 
    A_read <= 0; 
    B_read <= 0; 
    reg_A <= A_data_in; 
    reg_B <= B_data_in; 
    reg_C <= reg_C + alpha * reg_A * reg_B;     
    k <= k + 1; 
    if (k == N) begin
        k <= 0; 
        state <= WRITE_C; 
    end 
    else begin 
        state <= LOAD_A; 
    end 
end
...
\end{lstlisting}
\vspace{-10pt}
\caption{Feedback2}
\label{fig:feedback_loop2}
\end{subfigure}
\vspace{-5pt}
\caption{Inputs and error checking (loop)}
\label{fig:LLM_loop}
\end{figure}

\begin{figure}
\centering
\begin{subfigure}[b]{0.95\linewidth}
\begin{lstlisting}[]
Please write a tcl code to create a bram IP with properties like this :
-vendor xilinx.com  
-version 8.4 
-module_name blk_mem_gen_0 
-Write_Depth_A {4096} 
-Write_Width_A {32}
\end{lstlisting}
\vspace{-10pt}
\caption{Prompt}
\label{fig:prompt_TCL}
\end{subfigure}

\begin{subfigure}[b]{0.95\linewidth}
\begin{lstlisting}[language=TCL]
# Create BRAM IP with specific properties
create_ip -name blk_mem_gen -vendor xilinx.com -library ip -version 8.4 -module_name blk_mem_gen_0
# Set properties for BRAM IP
set_property -dict [list \
    CONFIG.Enable_A {Always_Enabled} \
    CONFIG.Register_PortA_Output_of_Memory_Primitives    
    {false} \
    CONFIG.Write_Depth_A {4096} \
    CONFIG.Write_Width_A {32} \
] [get_ips blk_mem_gen_0]

\end{lstlisting}
\vspace{-10pt}
\caption{Feedback}
\label{fig:feedback_TCL}
\end{subfigure}
\vspace{-5pt}
\caption{Inputs and error checking (IP)}
\label{fig:LLM_IP}
\end{figure}

\begin{figure}
\centering
\begin{subfigure}[b]{0.95\linewidth}
\begin{lstlisting}[]
Please write a testbench for this Verilog code, with instantiated BRAM. Use the 32-bit IEEE-754 format and set all initial values of the inputs to be between 0 and 1, where N = 64 and the period is 10ns. The float_to_ieee754 function is provided. Stop when done, and use fixed input values do not use random values. For input A, set A[i][j] = ((i + j) % 100) / 100.0; and x[i] = (i % 100) / 100.0. The initial integer should be declared outside the initial block.
\end{lstlisting}
\vspace{-10pt}
\caption{Prompt}
\label{fig:prompt_Testbench}
\end{subfigure}

\begin{subfigure}[b]{0.95\linewidth}
\begin{lstlisting}[language=verilog]
...
// Instantiate the syrk module
syrk #(.N(N), .DATA_WIDTH(DATA_WIDTH)) syrk_inst (
    .clk(clk),
    .rst(rst),
    .start_computation(start_computation),
    .alpha(alpha),
    .beta(beta),
    .A_data_in(A_data_in),
    .B_data_in(B_data_in),
    .C_data_out(C_data_out),
    .A_addr_out(A_addr_out),
    .B_addr_out(B_addr_out),
    .C_addr_out(C_addr_out),
    .A_read(A_read),
    .B_read(B_read),
    .C_write(C_write),
    .done(done)
);
...
\end{lstlisting}
\vspace{-10pt}
\caption{Feedback}
\label{fig:feedback_Testbench}
\end{subfigure}
\vspace{-5pt}
\caption{Inputs and error checking (Testbench)}
\label{fig:LLM_Testbench}
\end{figure}

\subsection{LLM-Assisted HLS Approaches}
Here, we describe the three LLM-assisted approaches explored herein, showcasing the diverse ways in which LLMs can contribute to hardware design. The direct LLM translation approach, denoted as \textit{C$\rightarrow$LLM$\rightarrow$Verilog}, and the natural language to Verilog approach, denoted as \textit{NL$\rightarrow$LLM$\rightarrow$Verilog}, demonstrate the capability of LLMs to generate Verilog directly from either code or natural language descriptions. The natural language to benchmark approach, denoted as \textit{NL$\rightarrow$LLM$\rightarrow$HLS-C}, on the other hand, highlights the potential for LLMs to augment existing HLS tools by raising the level of abstraction to natural language input. Figure \ref{fig:LLM_method} illustrates the design flow for each of these LLM-assisted HLS methodologies.

\subsubsection{C$\rightarrow$LLM$\rightarrow$Verilog}
The use of LLMs to directly generate synthesizable hardware accelerators in Verilog requires a well-defined procedure. This procedure involves the steps to generate Verilog code from high-level specifications and subsequent steps to produce a fully functional accelerator, from simulation to place-and-route. For example, a testbench is necessary to validate the accelerator's functionality during simulation. A place-and-route-ready hardware accelerator consists of Verilog code, TCL commands to automate the assembly of the accelerator's design (instantiating IP cores, connecting them, and setting up the overall project structure), and XDC files to specify the constraints of the accelerator such as clock period and I/O delay.

Figures \ref{fig:LLM_background}, \ref{fig:LLM_IO}, \ref{fig:LLM_loop}, \ref{fig:LLM_IP}, and \ref{fig:LLM_Testbench} illustrate our \textit{C$\rightarrow$LLM$\rightarrow$Verilog} process for different components of the hardware design flow. The first step defines the context of the generation process, including, but not limited to, the designer's role, the hardware background, and the constraints that the LLM (ChatGPT-4o, in our case) should follow to better identify the corresponding context and purpose of this process. Figure \ref{fig:LLM_background} shows the context we used in our experiments. We identify ourselves as hardware engineers and aim to translate a C program to HDL in Verilog. We specify that this Verilog module should target the Xilinx FPGA part xc7a200tfbg-484-1. Although ChatGPT-4o records the part in its memory, the design is not guaranteed to meet the I/O or resource constraints unless we explicitly instruct the LLM to meet the I/O constraints. If the specification of the part does not exist or is incorrect in the LLM, we must manually provide this information to the LLM.

After providing the role, background, and constraints of the designer and hardware to the LLM, we provide the source code to the LLM. It is important to be mindful of ChatGPT-4o's limitations: a 128k token limit for combined input and output, with a maximum of 4k tokens for the output alone. If a larger program is needed, it should be divided accordingly. In our experiments, all C benchmarks were within the 128k token limit, allowing us to input the entire program at once. However, due to the 4k output constraint, generating the complete Verilog accelerator required multiple iterations. Once generated, the Verilog output undergoes syntax and design error checking.

For designers proficient in hardware design, syntax and design error checking can be performed directly within the LLM. Otherwise, a validation tool like Vivado is necessary. Once an error is identified, we describe the error in natural language to the LLM and regenerate the Verilog code. This process is repeated until successful simulation and implementation in Vivado. We encountered some common errors in the process, such as incorrect data type mapping in I/O (Figure \ref{fig:LLM_IO}), misrepresentation of sequential and parallel execution (Figure \ref{fig:LLM_loop}), and state machine implementation errors (figures omitted for brevity). The designer's expertise level significantly impacts the speed and efficiency of this iterative error resolution process.

The final step in the LLM-assisted design flow is generating TCL scripts for IP integration, XDC constraints, and testbench content (Figures \ref{fig:LLM_IP} and \ref{fig:LLM_Testbench}). This step faces similar challenges as previous steps if the LLM lacks knowledge of the latest syntax or specifications, leading to more errors in generated files. For example, defining a proper clock period and calculating IEEE 754 standard floating-point values require the latest specifications. To address this problem, we manually provided the necessary information to the LLM, which learns and adapts over time, potentially reducing errors in future iterations.

\subsubsection{NL$\rightarrow$LLM$\rightarrow$Verilog}
The second approach is similar to \textit{C$\rightarrow$}\textit{ LLM$\rightarrow$Verilog} but uses natural language descriptions (or pseudocode) of the program's functionality as input to the LLM, instead of a programming language like C/C++. We described details such as input/output, variable types, loops, and operations. The number of prompts required in this approach depends on the complexity of the program and the designer's preferences, with LLMs like ChatGPT-4o potentially accommodating the entire program in a single prompt, as in our experiments.

\subsubsection{NL$\rightarrow$LLM$\rightarrow$HLS-C}
The third approach differs from the previous two by leveraging the strengths of both LLMs and traditional HLS tools. Instead of generating Verilog directly, it utilizes an LLM to translate natural language descriptions into HLS-compatible input (HLS-C), which is then processed by the HLS tool to produce the synthesizable Verilog output. This approach combines the expressiveness of natural language with the power and completeness of existing HLS tools, ultimately lowering the barrier to entry for hardware design by minimizing the need for proficiency in high-level programming languages.

\section{Experimental Setup} \label{Sec:exp}
To evaluate the three LLM-based approaches and compare them with the baseline HLS approach, we used nine benchmarks (\textit{syrk, syr2k, mvt, k3mm, k2mm, gesummv, gemm, bicg,} and \textit{atax}) from the Polybench suite \cite{Polybench}, specifically designed for evaluating the performance of HLS tools and compiler technologies. These benchmarks encompass computational kernels common in scientific and engineering applications, such as matrix multiplication, 2D convolution, and Cholesky decomposition. We employed ChatGPT-4o as our LLM model, Vitis HLS 2023.2 as our HLS tool, and Vivado 2023.2 for implementation targeting a Xilinx xc7a200tfbg484-1 FPGA. For each benchmark, we generated designs using all four approaches and collected data on resource utilization, power consumption, execution cycles, and critical path delay from Vitis HLS and Vivado. Note that the \textit{NL$\rightarrow$LLM$\rightarrow$Verilog} approach yielded an initial Verilog design with an equivalent structure to the initial Verilog design generated using the \textit{C$\rightarrow$LLM$\rightarrow$Verilog} approach. As such, these approaches share the same steps after the initial input stage, and thus have the same evaluation data. We tracked the number of prompts used to generate HLS-C, Verilog, TCL, XDC, and testbench content for the LLM-based approaches. For a fair comparison, we disabled automatic optimizations like pipelining in Vitis HLS. For LLM-based approaches, we used LLMs to generate all necessary content (Verilog code, TCL scripts, IPs, testbenches, XDC files) to form a complete project.

\begin{table}[t]
    \caption{The number of Prompts for LLM-based approaches}
    \vspace{-5pt}
    \label{tab:Prompts}
    \centering
    \begin{tabular}{cccccc}
    \hline
    \textbf{Benchmark} & \textbf{HLS-C} & \textbf{Verilog}  & \textbf{TCL}  & \textbf{Testbench} & \textbf{XDC}\\ 
    \hline
    syrk & 4 & 50 & 9 & 12 & 5\\ 
    \hline
    syr2k & 1 & 20 & 3 & 7 & 3\\ 
    \hline
    mvt & 1 & 36 & 3 & 7 & 2\\ 
    \hline
    k3mm & 1 & 21 & 3 & 5 & 2\\ 
    \hline
    k2mm & 1 & 29 & 3 & 6 & 3\\ 
    \hline
    gesummv & 1 & 23 & 3 & 7 & 3\\ 
    \hline
    gemm & 1 & 22 & 3 & 6 & 3\\
    \hline
    bicg & 1 & 16 & 3 & 6 & 3\\ 
    \hline
    atax & 1 & 11 & 3 & 8 & 3\\ 
    \hline
    \end{tabular}
\end{table}

\begin{table*}[t]
    \caption{Place \& routing results for \textit{C$\rightarrow$LLM$\rightarrow$Verilog} and \textit{C$\rightarrow$HLS$\rightarrow$Verilog} approaches}
    \label{tab:Result}
    \vspace{-5pt}
    \centering
    \begin{tabular}{cccccccccc}
    \hline
    \textbf{Benchmark} & \textbf{Approach} & \textbf{Execution cycles}  & \textbf{FF}  & \textbf{LUT} & \textbf{Slice} & \textbf{DSP} & \textbf{BRAM} & \textbf{Power (W)} & \textbf{CP}\\ 
    \hline
    \multirow{2}{*}{syrk} & LLM & 1859983 & 954 & 5300 & 1649 & 6 & 8 & 0.164 & 9.934\\ 

    & HLS & 3744260 & 662 & 521 & 221 & 5 & 32 & 0.350 & 8.191\\ 

    \hline
    \multirow{2}{*}{syr2k} & LLM & 2125846 & 542 & 472 & 197 & 2 & 12 & 0.181 & 9.446\\ 

    & HLS & 9028229 & 1042 & 960 & 1649 & 5 & 56 & 0.383 & 6.872\\ 

    \hline
    \multirow{2}{*}{mvt} & LLM & 44996 & 8628 & 2663 & 4404 & 2 & 4 & 0.197 & 9.312\\ 

    & HLS & 119492 & 713 & 991 & 342 & 5 & 12 & 0.332 & 6.55\\ 
    \hline
    \multirow{2}{*}{k3mm} & LLM & 2371593 & 623 & 328 & 236 & 2 & 28 & 0.207 & 9.924\\ 

    & HLS & 10277509 & 927 & 956 & 1649 & 5 & 56 & 0.398 & 6.646\\ 

    \hline
    \multirow{2}{*}{k2mm} & LLM & 1863816 & 537 & 311 & 202 & 2 & 20 & 0.189 & 9.967\\ 

    & HLS & 7963269 & 929 & 659 & 313 & 5 & 56 & 0.400 & 6.814\\ 
    \hline
    \multirow{2}{*}{gesummv} & LLM & 65991 & 437 & 288 & 170 & 2 & 16 & 0.176 & 9.253\\ 

    & HLS & 148805 & 795 & 561 & 228 & 5 & 20 & 0.316 & 6.855\\ 
    \hline
    \multirow{2}{*}{gemm} &LLM & 1601739 & 488 & 332 & 200 & 2 & 16 & 0.178 & 9.697\\ 

    & HLS & 4542980 & 807 & 505 & 238 & 5 & 32 & 0.359 & 6.551\\ 
    \hline
    \multirow{2}{*}{bicg} & LLM & 46478 & 505 & 194 & 198 & 2 & 20 & 0.196 & 9.251\\ 

    & HLS & 119492 & 711 & 429 & 223 & 5 & 12 & 0.333 & 6.599\\ 
    \hline
    \multirow{2}{*}{atax} & LLM & 57669 & 453 & 257 & 164 & 2 & 16 & 0.167 & 9.952\\ 

    & HLS & 119492 & 741 & 428 & 209 & 5 & 11 & 0.309 & 6.573\\ 
    \hline
    \end{tabular}
\end{table*}

\section{Results and Analysis}
Table \ref{tab:Prompts} presents the number of prompts required for each file type (HLS-C, Verilog, TCL, testbench, and XDC) to construct a complete hardware accelerator from C benchmarks. As demonstrated in Sec. \ref{Sec:exp}, the \textit{C$\rightarrow$LLM$\rightarrow$Verilog} and \textit{NL$\rightarrow$LLM$\rightarrow$Verilog} approaches share the same prompts after the initial input, leading to identical place-and-route results. For the \textit{NL$\rightarrow$LLM$\rightarrow$HLS-C} approach, we also include the number of prompts needed to generate the HLS-C code. Since we targeted the same functionality as the C benchmark,  the \textit{NL$\rightarrow$LLM$\rightarrow$HLS-C} and \textit{C$\rightarrow$HLS$\rightarrow$Verilog} approaches share the same place-and-route outcomes. 

Notably, generating the Verilog code generally required the most prompts compared to other file types. But the number of prompts required varied significantly depending on the benchmark, as well as our growing familiarity with the LLM's behavior with Verilog generation. The \textit{syrk} benchmark, for example, required considerably more interaction with the LLM compared to \textit{atax} (the last benchmark we worked on). The \textit{syrk} kernel exhibits a higher level of complexity, containing four nested loops with multiple multiplications in a single operation and three 2D arrays for inputs and outputs. Conversely, \textit{atax} only comprises two nested loops and one 2D array for input. This suggests that the inherent complexity of the benchmark code, as well as our initial learning curve to effectively prompt the LLM to minimize errors, heavily influenced the number of prompts needed for accurate Verilog generation. As we gained experience and refined our prompting strategies, we were able to consolidate prompts, leading to faster generation for subsequent benchmarks. In contrast, the number of prompts for TCL generation remained relatively consistent across all benchmarks, implying that this task is less sensitive to the specific characteristics of the input code. The complexity of the benchmark and the designer's growing familiarity with LLM interaction are key factors in determining the number of prompts needed for successful Verilog generation, although prior design experience can also play a role.

Table \ref{tab:Result} presents the simulation and implementation results for both LLM-based and HLS-based approaches. For each benchmark, LLM refers to the \textit{C$\rightarrow$LLM$\rightarrow$Verilog} and \textit{NL$\rightarrow$LLM$\rightarrow$Verilog} approaches, while HLS refers to \textit{NL$\rightarrow$LLM$\rightarrow$HLS-C} and \textit{C$\rightarrow$HLS$\rightarrow$Verilog} approaches. To determine the quality of a resulting hardware accelerators, the evaluation metrics include execution cycles, resource utilization (FFs, LUTs, Slices, DSPs, and BRAMs), total power consumption, and critical path delay.

A key observation is the significant variation in results across different benchmarks. For the \textit{syrk} and \textit{mvt} benchmarks, the LLM-based approaches consume more resources (except DSPs and BRAMs) compared to HLS. This is attributed to the use of LUT RAM for the inner matrix in the LLM-generated designs. 

However, for the remaining seven benchmarks, LLM-based approaches consistently outperformed the HLS-based approaches across all metrics. This includes a notable reduction in resource utilization (with an average decrease of 38.67\%), a significant improvement in execution cycles (average reduction of 64\%), and a substantial reduction in total power consumption (average reduction of 38.67\%). For the critical path, the HLS-based approach outperformed LLM-based approaches by an average of 28.82\%.

Overall, the results in Table \ref{tab:Result} demonstrate the potential of LLMs in optimizing various aspects of hardware design. While the LLM-based approaches did not outperform in every metric for all benchmarks, their consistent success in the majority of cases, particularly in resource utilization, power consumption, and often execution cycles, highlights the promise of this technology for HLS. Further research is needed to refine and expand these capabilities, and explore them in a wider variety of usage scenarios, but the current results are encouraging and suggest that LLMs could play a significant role in the future of hardware design automation.

\section {The Energy Elephant in the LLM-HLS Room}
While the initial excitement surrounding the integration of LLMs into the HLS workflow has spurred significant research, a critical aspect has been conspicuously absent from most discussions: the energy implications. The majority of studies have focused on the potential of LLMs to streamline the design process, enhance automation, and improve the quality of generated hardware. However, they have largely overlooked the energy consumption associated with both the training and inference of these models.

LLMs, particularly large-scale models like GPT-3 and GPT-4, are notorious for their computational demands. Training LLMs can consume hundreds of megawatt-hours to several gigawatt-hours of electricity \cite{schwartz2020greenAI}. Even inference, the process of generating responses to prompts, can be computationally intensive, requiring substantial energy resources. The Electrical Power Research Institute (EPRI) estimates that a single ChatGPT query can consume approximately 2.9 W-hours of energy---nearly 10 times the power of a single Google search \cite{EPRI2024}---a considerable amount when numerous queries are needed for HLS tasks. This raises concerns about the overall energy efficiency of incorporating LLMs into the HLS flow. Given that a primary goal of HLS is to design hardware accelerators that are more energy efficient than general-purpose computers, the energy overhead of utilizing LLMs could outweigh the intended benefits.

Furthermore, the process of fine-tuning LLMs for specific HLS tasks can exacerbate the issue of energy consumption. Fine-tuning involves retraining the model on domain-specific data, which is computationally expensive. If the energy cost of fine-tuning and utilizing an LLM is greater than the energy saved across all resulting hardware designs, then employing LLMs in this way would be counterproductive for energy efficiency.

The lack of attention to power/energy implications in current research raises concerns about the sustainability and practicality of LLM-driven HLS. As the field progresses, it is imperative to thoroughly investigate and quantify the energy costs associated with LLM utilization. This will enable a more comprehensive evaluation of the trade-offs between design efficiency and power consumption, ultimately leading to more informed decisions regarding the appropriate use of LLMs in HLS.

\section{Conclusion} \label{sec:concl}

This paper has explored the application of Large Language Models (LLMs) in High-Level Synthesis (HLS), evaluating their potential to transform hardware design workflows. Through a survey and experimental evaluations, we assessed the ability of LLMs to generate Verilog code from high-level specifications, including both C benchmarks and natural language descriptions. Our findings reveal that LLM-based approaches can significantly enhance the efficiency of the HLS process, demonstrating notable improvements in resource utilization, execution cycles, and power consumption for most benchmarks compared to traditional HLS tools. However, challenges remain in ensuring the quality and optimization of LLM-generated code, particularly regarding critical path delays and the complexity of initial prompt interactions. Additionally, the substantial energy consumption associated with training and utilizing LLMs raises concerns about the overall energy efficiency of their integration into HLS workflows. Despite these challenges, the promising results suggest that with further refinement and research, LLMs could play a pivotal role in the future of hardware design automation, offering a powerful tool to streamline and optimize the HLS process.

\begin{acks}
This work was partially supported by the Technology and Research Initiative Fund (TRIF) provided to the University of Arizona by the Arizona Board of Regents (ABOR) and by NSF Grant 1844952.
\end{acks}

\bibliographystyle{ACM-Reference-Format}
{\small
\bibliography{refs}}
\balance

\end{document}